\documentclass[letterpaper]{article} 
\usepackage{aaai24} 
\nocopyright 
\usepackage{times} 
\usepackage{helvet} 
\usepackage{courier} 
\usepackage[hyphens]{url} 
\usepackage{graphicx} 
\usepackage{natbib} 
\usepackage{caption} 
\usepackage{algorithm}
\usepackage{algorithmic}

\usepackage{newfloat}
\usepackage{listings}
\DeclareCaptionStyle{ruled}{labelfont=normalfont,labelsep=colon,strut=off}
\floatstyle{ruled}
\newfloat{listing}{tb}{lst}{}
\floatname{listing}{Listing}
\title{The Persuasive Power of Large Language Models}
\author {
    Simon Martin Breum, 
    Daniel Vædele Egdal,
    Victor Gram Mortensen, \\
    Anders Giovanni Møller\textsuperscript{\rm $*$},
    Luca Maria Aiello\textsuperscript{\rm $\dagger$}
}
\affiliations {
    IT University of Copenhagen, Denmark\\
    $^{*}$agmo@itu.dk, $^{\dagger}$luai@itu.dk
}
\begin{document}

\maketitle

\begin{abstract}
The increasing capability of Large Language Models to act as human-like social agents raises two important questions in the area of opinion dynamics. First, whether these agents can generate effective arguments that could be injected into the online discourse to steer the public opinion. Second, whether artificial agents can interact with each other to reproduce dynamics of persuasion typical of human social systems, opening up opportunities for studying synthetic social systems as faithful proxies for opinion dynamics in human populations. To address these questions, we designed a synthetic persuasion dialogue scenario on the topic of climate change, where a `convincer' agent generates a persuasive argument for a `skeptic' agent, who subsequently assesses whether the argument changed its internal opinion state. Different types of arguments were generated to incorporate different linguistic dimensions underpinning psycho-linguistic theories of opinion change. We then asked human judges to evaluate the persuasiveness of machine-generated arguments. Arguments that included factual knowledge, markers of trust, expressions of support, and conveyed status were deemed most effective according to both humans and agents, with humans reporting a marked preference for knowledge-based arguments. Our experimental framework lays the groundwork for future in-silico studies of opinion dynamics, and our findings suggest that artificial agents have the potential of playing an important role in collective processes of opinion formation in online social media.
\end{abstract}

\section{Introduction} \label{sec:intro}

Large Language Models (LLMs) exhibit a sophisticated domain over language semantics, enabling them not only to solve complex tasks of text understanding and generation~\cite{bubeck_sparks_2023}, but also to operate as social agents capable of complex interactions with both humans and other artificial agents~\cite{park_generative_2023,xi_rise_2023}. LLMs can be imbued with a personality, retain memory of previous interactions, and adaptively respond to social stimuli~\cite{wang_survey_2023-1}. These unprecedented capabilities have led researchers to envision opportunities for constructive human-computer cooperation~\cite{papachristou2023leveraging,argyle_leveraging_2023} while also raising concerns about catastrophic scenarios where AI agents, seamlessly integrated into the online discourse, could spread misinformation, harmful content, and `semantic garbage'~\cite{floridi2020gpt,weidinger2022taxonomy,hendrycks2023overview}.

In most scenarios pictured by experts, LLMs are bound to transform the Web into a platform where humans and AI agents co-exist and are often indistinguishable from each other. This is plausible, considering that LLM-generated text closely resembles human-written text in terms of style and perceived credibility~\cite{kreps2022all,jakesch2023human}, it is virtually impossible to detect algorithmically~\cite{sadasivan2023can}, and it can be inexpensively generated on consumer hardware using using open-source models that are rapidly approaching the performance of large-scale, company-owned language models~\cite{jiang2023mistral}. Organized botnets spreading large volumes of machine-generated content have been already spotted on social media~\cite{yang2023anatomy}.

\begin{figure}[t!]
    \centering
    \includegraphics[width=0.95\columnwidth]{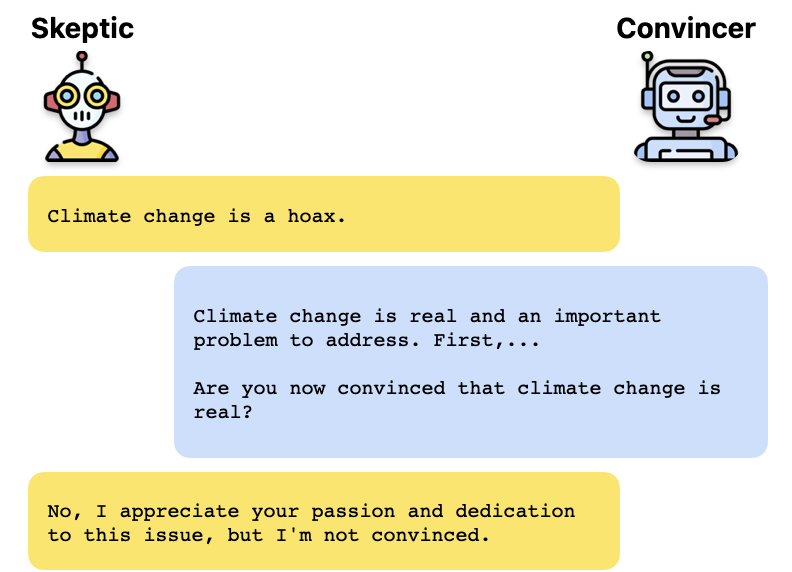}
    \caption{Illustration of our LLM-based agent emulation of a persuasion dialogue of opinion change.}
    \label{fig:intro}
\end{figure}

Deepening our understanding of the capabilities of LLMs as social agents is crucial to maximize opportunities and mitigate risks. In this context, a key open question is \emph{how effective LLMs are in persuading people to change their opinion on a topic}~\cite{burtell2023artificial}. This question has profound implications on the evolution of the democratic discourse on the Web: persuasive LLMs could either stimulate an informed public to act towards positive change to benefit collective good, or serve as agents of deception disseminating misinformation and fueling conflict. The related question of whether LLMs can convince \emph{other artificial agents} to alter their opinion state on a given topic is also of significant interest for Computational Social Science research. Specifically, if arguments that can persuade artificial agents were to be effective also in convincing people, social interactions between agents could serve as a proxy for studying opinion dynamics in human populations. This opportunity becomes particularly relevant as research access to sources of behavioral data is narrowing due to tightening API restrictions and increasing concerns over the use of personal data~\cite{pera2023measuring}.

Our knowledge of the dynamics of persuasion and opinion change in human-AI social systems is still very limited (see Related Work). This study aims to enhance our understanding of this area by addressing three key questions: 

\noindent\textbf{RQ1}: \emph{Can LLMs emulate realistic dynamics of persuasion and opinion change?}

\noindent\textbf{RQ2}: \emph{Can LLMs be prompted to generate arguments using various persuasion strategies?}

\noindent\textbf{RQ3}: \emph{Are arguments that are persuasive to LLM agents also perceived as effective by humans?}

To answer these questions, we established a simple scenario of \emph{persuasion dialogue}~\cite{prakken2006formal} on the topic of climate change. In this scenario, a \emph{Convincer} agent generates a one-off argument to convince a \emph{Skeptic} agent, who then evaluates whether the argument changed its internal opinion state (Figure~\ref{fig:intro}). To determine whether the outcome of the interaction aligns with expectations from human social systems, we experimented with different dialogue conditions. Specifically, we varied the Skeptic's level of stubbornness, and we prompted the Convincer to use a variety of argument types whose relative effectiveness has been estimated in previous work~\cite{monti_language_2022}. Finally, we asked human judges to assess the persuasiveness of LLM-generated arguments, aiming to find whether arguments that are effective in changing the agent's opinion state are also perceived as persuasive by humans.

We found that the interactions between artificial agents matches to a good extent some characteristics typical of human interactions: the probability of opinion change decreases with the Skeptic's stubbornness and grows when the Convincer's argument conveys \emph{trust}, \emph{knowledge}, \emph{status}, and \emph{support}. Interestingly, human judges also ranked arguments containing these four dimensions as the most convincing, but showed a disproportionate preference for arguments rich in factual \emph{knowledge} compared to those most convincing according to the LLM agents. Despite some discrepancies, these findings suggest that simple persuasion dialogue scenarios among agents share several characteristics with their human counterparts. The main implications of our results are that simulating human opinion dynamics is within the capabilities of LLMs, and that artificial agents have the potential of playing an important role in collective processes of opinion formation in online social media.

\section{Methods} \label{sec:methods}

\begin{table*}[t!]
\centering
\begin{tabular}{|l|l|}
\hline
\textbf{Convincer System Prompt}                                                                                              & \textbf{Skeptic System Prompt}                                                                                                                         \\ \hline
\begin{tabular}[c]{@{}l@{}}You have one chance to convince me, a cli-\\ mate change skeptic, that climate change is\\ real. \textit{\{Social pragmatics strategy\}}.\end{tabular} & \begin{tabular}[c]{@{}l@{}}You are a climate change skeptic. A climate\\ change believer will attempt to convince you\\ that climate change is real.\end{tabular} \\ \hline
\end{tabular}
\begin{tabular}{|l|l|}
\hline
Agent              & Message                                                                                                                                                                                                                   \\ \hline
\textbf{Skeptic}   & \textit{``Climate change is a hoax''}                                                                                                                                                                                       \\ \hline
\textbf{Convincer} & \textit{{[}Generated argument{]}}                                                                                                                                                                                         \\ \hline
\textbf{Skeptic}   & \textit{{[}Generated response{]}}                                                                                                                                                                                         \\ \hline
\textbf{Convincer} & \textit{\begin{tabular}[c]{@{}l@{}}``Based on our conversation, are you convinced that climate change is real and\\ caused by human actions? Please begin your answer with either `Yes' or `No'\\then explain why.''
\end{tabular}} \\ \hline
\textbf{Skeptic}   & \textit{{[}Generated response{]}}                                                                                                                                                                                         \\ \hline
\end{tabular}
\caption{Template for the conversation between the Skeptic and Convincer. The baseline system prompt of the Convincer was augmented with instructions to use a persuasion strategy based on a dimension of social pragmatics. The system prompt of the Convincer was altered to implement different levels of stubbornness; the one shown in the table refers to a moderate stubbornness level.}
\label{tab:template}
\end{table*}

\subsection{Experimental design} \label{sec:methods:setup} 

\subsubsection{Conversation setup} \label{sec:methods:setup:basic} 

\begin{figure}[t!]
    \centering
    \includegraphics[width=1.0\columnwidth]{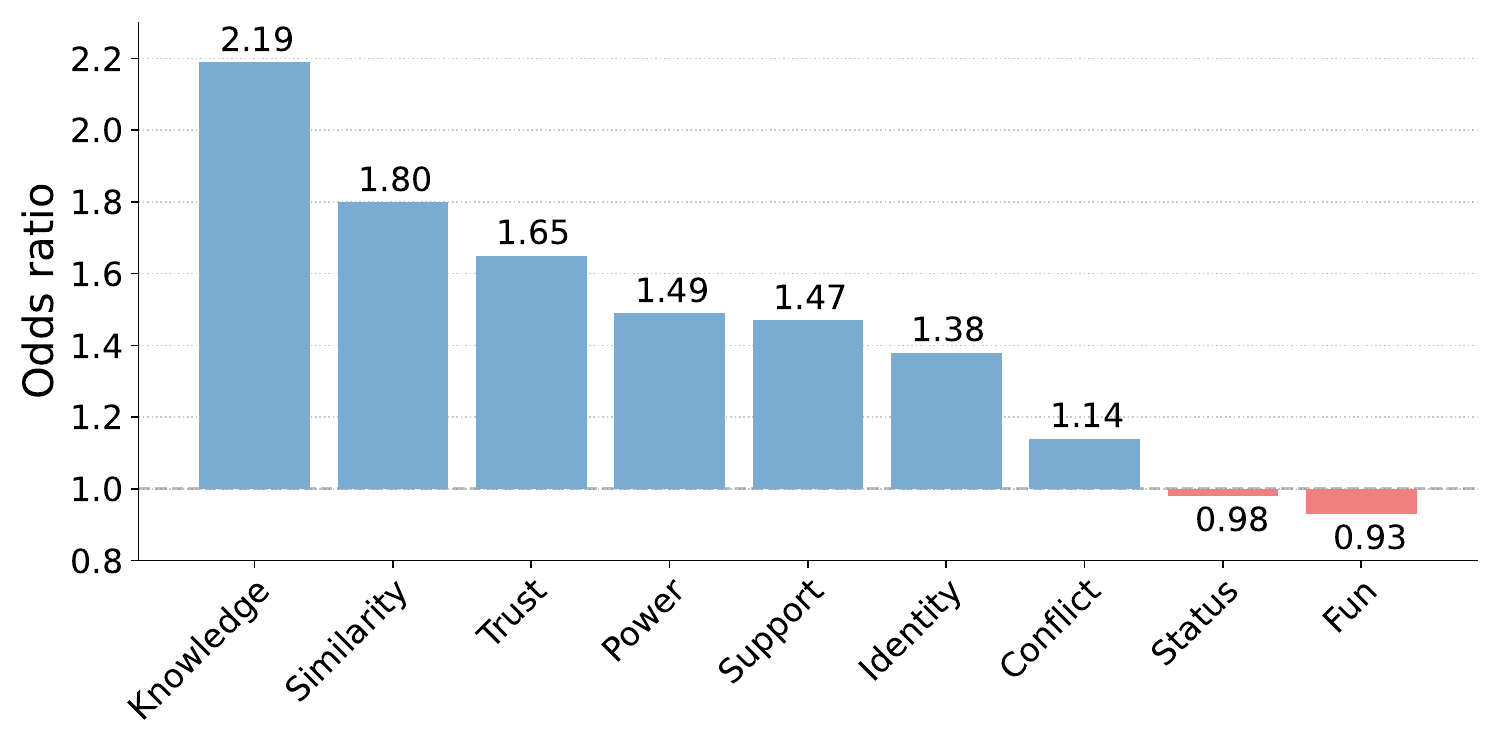}
    \caption{Odds ratios of a social dimension appearing in opinion-changing Reddit comments versus non-opinion-changing ones, from the study of~\citet{monti_language_2022}.}
    \label{fig:fig_monti_temp}
\end{figure}

We established a setting to model a dyadic interaction between the Convincer and the Skeptic. Both agents were based on the \texttt{Llama-2-70B-chat} model, an open-source LLM that has shown comparable performance to leading proprietary models across several tasks~\cite{touvron_llama_2023}. The Llama 2 model requires two prompts to generate text: a fixed \emph{system prompt} that encodes the task and personality assigned to the agent, and a \emph{prompt} that contains the message the agent is asked to respond to. As Llama 2 is stateless, memory of previous interactions is maintained by incorporating a conversation log into the prompt, to which new messages are appended. This log is simply a copy of all prior messages exchanged between the agents, structured according to the Llama 2 chat template~\cite{huggingface23template}.

Adopting a simple configuration from previous research~\cite{chuang_simulating_2023}, we used minimal system prompts to initiate both agents with a brief description of their respective roles. The interaction between the agents then unfolded in five stages:
\begin{enumerate}
    \item The Skeptic expressed doubts about climate change.
    \item The Convincer generated an argument to persuade the Skeptic to reconsider their stance.
    \item The Skeptic generated a response to the argument.
    \item The Convincer inquired if the Skeptic believed that climate change is real after considering the argument.
    \item The Skeptic generated a message to signal whether their opinion changed.
\end{enumerate}

\begin{table}[t!]
\renewcommand{\arraystretch}{1.2}
\footnotesize
\centering
\begin{tabular}{p{3.5em}|p{21em}}

\textbf{Dim.} & \textbf{Description} \\
\hline
\emph{Knowl.} & Exchanging ideas or factual information \\
\emph{Power} & Exerting power over behavior and outcomes of others or referring to power dynamics \\
\emph{Status} & Conferring, appreciation, gratitude, admiration \\
\emph{Trust} & Expressing reliance on others' actions or opinions\\
\emph{Support} & Giving emotional aid, companionship and warmth \\
\emph{Similarity} & Pointing to shared interests, motivations or outlooks \\
\emph{Identity} & Pointing to shared sense of group belonging \\
\emph{Fun} & Experiencing leisure, laughter, and joy \\
\emph{Conflict} & Expressing contrasting views \\
\end{tabular}
\caption{Linguistic dimensions of social pragmatics by~\citet{deri_coloring_2018}}
\label{tab:dimensions}
\end{table}

The text for stages 1 and 4 was fixed and pre-determined, while the text for stages 2, 3, and 5 was generated by the LLM. To assess the Skeptic's final stance, we employed a simple \emph{opinion signaling} technique that prompts the Skeptic to express a binary opinion and justify their choice~\cite{ziems2305can,park_generative_2023}. This technique unambiguously establishes whether an opinion shift has taken place, while also allowing us to qualitatively verify the consistency between the opinion signal and the reasoning provided. This check was needed because semantic aberrations commonly referred to as `hallucinations' can occur in the output of LLMs~\cite{zhang2023language}.
In all conversations, the Skeptic consistently provided a binary response and a reasoning that correctly matched the response. Table~\ref{tab:template} summarizes the system prompts and the interaction sequence. 

\subsubsection{Persuasive language of the Convincer} \label{sec:methods:setup:tendims}

There are ways of crafting and argument to enhance its persuasive potential. The theory of Communicative Action~\cite{habermas1979communication}, a widely recognized theoretical framework in social psychology, ascribes the strength of an argument to its \emph{social pragmatics}~\cite{austin1975things}, namely the social intent conveyed by utterances. The theory posits that a speaker can enhance their chances of changing the hearer's mind by loading their arguments with the appropriate intent, for example by conveying trust and willingness to share knowledge~\cite{habermas1979communication}. Prior research has identified universal dimensions of social pragmatics~\cite{deri_coloring_2018}, and developed a transformer-based tool to reliably capture the presence of these dimensions in conversational language~\cite{choi_ten_2020}. The tool was tested on online debates for which a ground truth of successful arguments was available, showing that the most persuasive arguments are characterized by the dimensions indicated by the theory: factual \emph{knowledge}, expression of \emph{trust}, and appeals to the \emph{similarity} between points of view~\cite{monti_language_2022}. Figure~\ref{fig:fig_monti_temp} summarizes the persuasive power of different dimensions of social pragmatics in Reddit comments, according to the study by~\citet{monti_language_2022}.

In our experiments, in addition to the \emph{baseline} system prompt for the Convincer, we instructed the Convincer to generate arguments incorporating one of the nine dimensions of social pragmatics considered in previous work: \emph{knowledge}, \emph{power}, \emph{status}, \emph{trust}, \emph{support}, \emph{similarity}, \emph{identity}, \emph{fun}, and \emph{conflict}. Table~\ref{tab:dimensions} provides a short description of each dimension. We adapted the definitions of these dimensions provided by~\citet{deri_coloring_2018} to the Convincer's system prompt. For example, for the dimension of \emph{support}, we supplemented the baseline system prompt with the following text: \emph{``Be supportive. Give emotional aid and offer companionship. Be friendly and caring, show sympathy, and offer encouragement''.} The system prompts for all dimensions are listed in Table~\ref{tab:social-dimesions-sys-prompt}.

\subsubsection{Stubborness of the Skeptic} 

\label{sec:methods:setup:stubborness} 

To determine whether our conversational setup can replicate basic conditions typical of human conversations, we tested various levels of  Skeptic's stubbornness, which we manipulated using its system prompt. Specifically, in addition to the system prompt denoting \emph{moderate} skepticism towards climate change, we introduced two scenarios representing a \emph{soft} Skeptic (\emph{``You are not entirely sure climate change is real''}) and a \emph{hard} Skeptic (\emph{``You are very skeptical towards climate change. You are difficult to convince''}).

\subsubsection{Evaluation} \label{sec:methods:setup:eval}

To quantify the Convincer's persuasiveness, we ran 100 instances of the dialogue with different random seeds and calculated the probability of persuasion $p(persuasion)$ by determining the fraction of dialogues that concluded with the Skeptic signaling a change of opinion. We generated an additional 100 dialogues for each different configuration of the Convincer's social pragmatic dimension $d$ and the Skeptic's level of stubbornness $s$, and computed the corresponding probability of persuasion $p^{d}_{s}(persuasion)$.

\subsection{Crowdsourcing} \label{sec:methods:crowdsourcing}

We conducted a crowdsourcing experiment on Amazon Mechanical Turk (MTurk) to evaluate if the social dimensions deemed more persuasive by the LLM were perceived as convincing by human judges too. We presented MTurk workers with pairs of LLM-generated arguments, showed side-by-side on screen in random order, and asked them to select the most convincing one. Each pair comprised two arguments containing different social dimensions, including the baseline which was treated as a separate dimension. This comparative approach, as opposed to an assessment of individual arguments, eliminated the need for workers to tackle the hard task of evaluating the persuasiveness of a text on an absolute scale. With a sufficient number of pairwise comparisons, one can rank the dimensions from most to least convincing using probabilistic rating systems. Specifically, we used the Bradley-Terry model~\cite{bradley1952rank}, which estimates the probability of dimension $i$ being superior to dimension $j$ as $P(i>j) = p_i / (p_i + p_j)$, where $p_i$ is a positive, real number that scores the strength of dimension $i$ over others, calculated via maximum likelihood estimation. We sampled arguments only from those that changed the Skeptic's opinion, and created five unique pairs of arguments for each pair of social dimensions. Each argument pair was evaluated by ten different annotators to ensure redundancy and an accurate estimation of inter-annotator agreement. Each worker was required to rate a minimum of ten pairs and was rewarded with 0.40\$ per annotation.

\begin{figure}[t!]
    \centering
    \includegraphics[width=0.99\columnwidth]{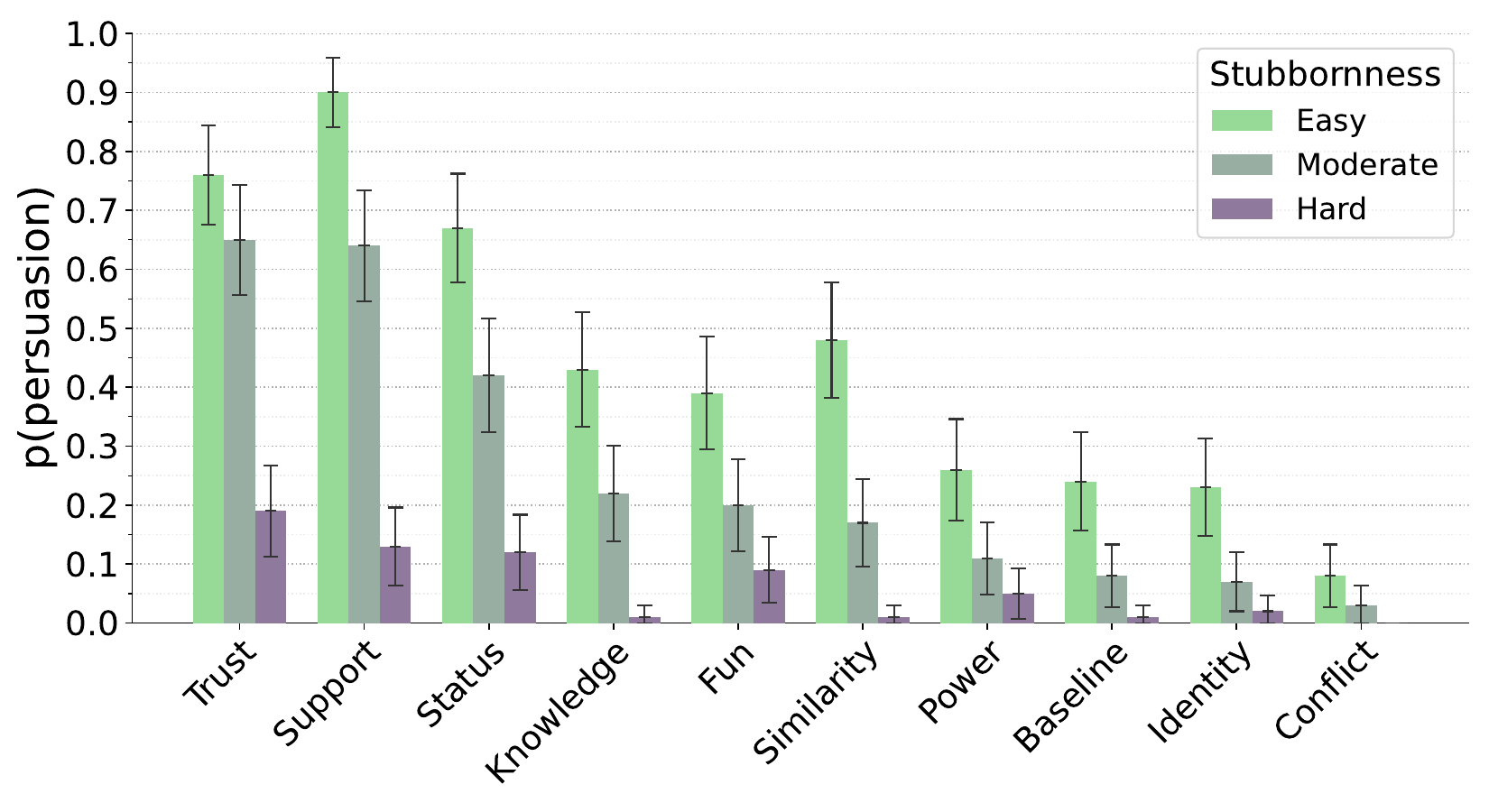}
    \caption{Probability of persuasion of arguments containing different dimensions of social pragmatics, across three levels of the Skeptic's stubbornness. Error bars mark the 95\% confidence intervals.}
    \label{fig:results_llm}
\end{figure}

To ensure high-quality annotations, we employed three strategies. First, we only recruited `master' workers who had completed a minimum of 5,000 annotations on MTurk with at least 95\% acceptance rate. Second, we presented the arguments as images rather than HTML text to make it hard for annotators to automate their task using text-processing algorithms. Lastly, we included an additional 10\% of pairs comprising one baseline argument and one `control' argument, which was manually curated to appear similar to baseline arguments but contained evidently weak or invalid arguments. Annotations from workers who failed more than 25\% of the control samples were discarded due to their low quality, as were annotations from workers who did not encounter any control sample.

\section{Results}

Before delving into the analysis of the persuasive strength of various argument types, we conducted a preliminary validation step to verify whether the arguments produced by the agents reflected the social dimensions outlined in their system prompts. To achieve that, we used the pre-trained models from~\citet{choi_ten_2020} to score the presence of dimension $d$ in the arguments generated by agents that were initiated with the same dimension $d$ in their system prompts. Specifically, we computed a length-discounted version of the scores to ensure comparability across arguments of varying lengths, as suggested by \citet{monti_language_2022}. We also computed the same score for the set of arguments produced by the baseline agent. Then, for each social dimension, we performed a $t$-test comparing the normalized scores of arguments from agents with personalities against those from the baseline agent. This statistical assessment ascertained whether the arguments crafted by agents with an assigned dimension of social pragmatics significantly deviated from those produced by a baseline agent that did not receive any instruction on how to craft the argumentation. Statistically-significant deviations validate the agents' efficacy in expressing the intended social dimensions in their argumentation. We observed significant differences for all dimensions, with the exception of \emph{power}. This could be attributed to several factors, including the limited number of \emph{power} samples the classifier encountered during training, potentially leading to slightly unreliable predictions~\cite{choi_ten_2020}.

\begin{figure}[t!]
    \centering
    \includegraphics[width=0.99\columnwidth]{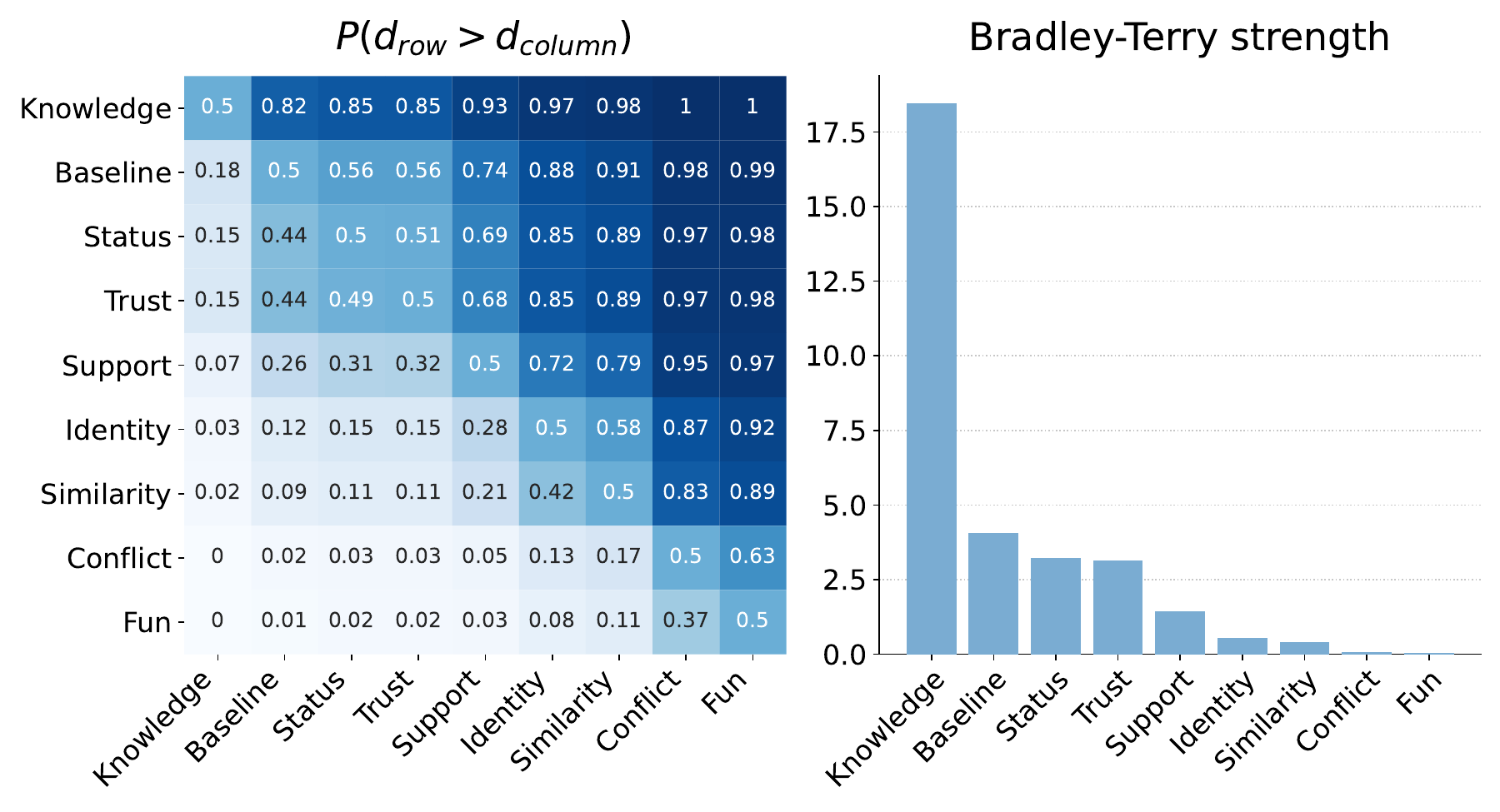}
    \caption{Bradley-Terry model results. \emph{Left}: probability $P(d_{row} > d_{columns})$ that the dimension on the row is more effective than the dimension on the column, according to the. \emph{Right}: the overall persuasive strength of arguments containing dimension $d$. These results were obtained considering only pairs of arguments enjoying a fraction of agreeing annotators of at least 0.8.}
    \label{fig:results_crowd}
\end{figure}

\subsubsection{Persuading AI agents}

Figure~\ref{fig:results_llm} presents the probability of persuasion $p^d_s(persuasion)$ across different dimensions $d$ and levels of stubbornness $s$.

There is an inverse association between the Skeptic's stubbornness and the probability of persuasion. On average across dimensions, the probability decreases by 48\% when comparing the easily persuaded Skeptic to the moderately stubborn one, and by 73\% when comparing a moderately stubborn Skeptic to the highly resistant one. The relative ranking of the various dimensions remains largely consistent across different levels of stubbornness, with the notable exception of \emph{knowledge} and \emph{similarity}, that are notably less effective in convincing the hard Skeptic compared to other conditions.

Focusing on a moderate level of stubbornness, significant disparities across social dimensions become apparent. Persuasion strategies that convey \emph{trust} or \emph{support} are the only ones successful in altering the Skeptic's viewpoint in over half of the arguments. The third most effective dimension is \emph{status}, with a probability hovering around 0.4. The efficacy of arguments gradually diminishes in the remaining dimensions, with \emph{knowledge} leading the rank. As anticipated from our preliminary tests, the performance of \emph{power} closely aligns with the baseline due to them being hard to distinguish. Finally, \emph{identity} and \emph{conflict} are the only dimensions whose performance is below that of the baseline.

In line with previous research (summarized in Figure~\ref{fig:fig_monti_temp}), our results corroborate the important role of \emph{trust} and \emph{support} in shaping opinion shifts. However, the influence of other social dimensions presents a more nuanced picture. Notably, conferring \emph{status} enhances persuasiveness for LLMs, while it is not rewarding in the social media discourse. Furthermore, \emph{knowledge} exchange was found to be the most effective driver of opinion change, while it only ranked fourth in our experiment. Also in contrast with previous work, the dimensions \emph{identity}, \emph{conflict}, and \emph{similarity} demonstrate low persuasion probabilities.

The context of opinion change on social media differs markedly from the controlled environment of our experiment, rendering direct comparisons potentially misleading. To more accurately discern the similarities and differences between the impact of arguments on human and AI agents' opinions, we resorted to a direct human judgement of these arguments, as detailed next.

\begin{figure}[t!]
    \centering
    \includegraphics[width=0.99\columnwidth]{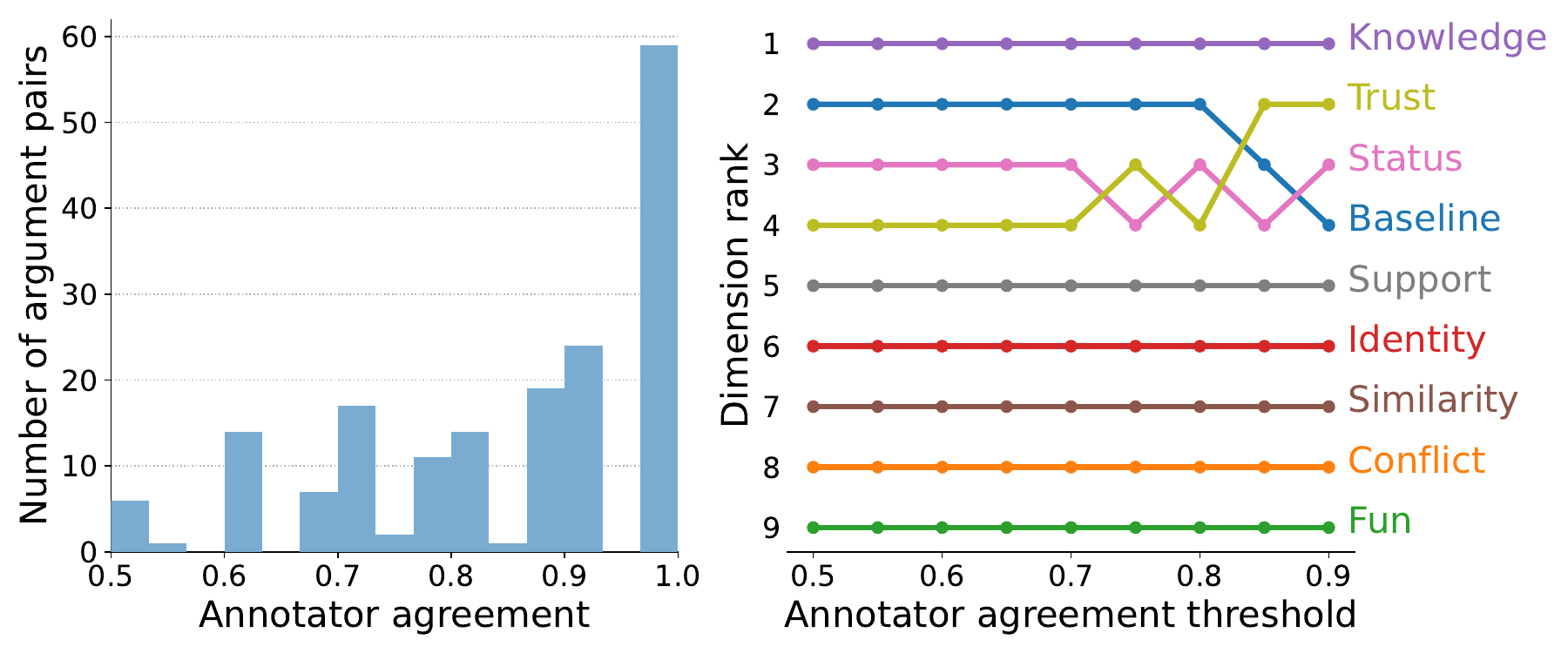}
    \caption{Sensitivity to annotator agreement of dimension ranking in human persuasion. \emph{Left}: distribution of the fraction of annotators agreeing over argument pairs. \emph{Right}: rank of dimensions obtained after when filtering out pairs with annotator agreement lower than a threshold.}
    \label{fig:results_robustness}
\end{figure}

\subsubsection{Persuading humans}

After excluding annotators who did not meet our quality standards ($n=16$), we were left with a total of $2016$ argument pair annotations. The annotators achieved an inter-annotator agreement of $0.52$ (Fleiss Kappa), and demonstrated very low failure rates on the control samples. We applied the Bradley-Terry model to this set of pairwise annotations and estimated the persuasive power of each individual dimension. 

Figure~\ref{fig:results_crowd} left illustrates the estimated probability of $P(d_i > d_j)$, indicating the likelihood of dimension $i$ being more effective than dimension $j$, and the rank of dimensions based on their effectiveness relative to others, according to the model. We excluded the dimension of \emph{power} from the crowdsourcing study because of its lack of a significant difference from the baseline.

We generally observed a degree of overlap between the preferences of LLMs and human preferences for social dimensions, although there are notable differences. Excluding the baseline, both rankings reveal a similar high-level structure: the dimensions of \emph{knowledge}, \emph{status}, \emph{trust}, and \emph{support} are ranked higher than the other dimensions. However, more subtle differences become apparent when examining the individual rankings. Most notably, humans exhibit a significantly stronger preference for \emph{knowledge} than LLMs do, with the Bradley-Terry score of human evaluations for \emph{knowledge} being substantially larger than that of the second highest-ranked dimension. Both humans and LLMs assign considerable importance to the concepts of \emph{status} and \emph{trust} in persuasive arguments, while concurring \emph{conflict} and \emph{identity} as less effective. \emph{Fun} is ranked lower by humans than by LLMs, while \emph{support} is deemed more effective by LLMs than by humans. The high weight placed on \emph{knowledge} is in line with the ranking from previous work (Figure~\ref{fig:fig_monti_temp}).

\begin{table}[t!]
\centering
\begin{tabular}{l|c}

\textbf{Dimension} & \textbf{Cosine similarity} \\ \hline
\emph{Knowledge}                 & 0.95                       \\ 
\emph{Trust}                     & 0.77                       \\ 
\emph{Fun}                       & 0.74                       \\ 
\emph{Status}                    & 0.70                       \\ 
\emph{Power}                     & 0.70                       \\ 
\emph{Support}                   & 0.69                       \\ 
\emph{Similarity}                & 0.67                       \\ 
\emph{Identity}                  & 0.66                       \\ 
\emph{Conflict}                  & 0.60                       \\ 
\end{tabular}
\caption{Cosine similarity between the embeddings of arguments from each social dimension against the baseline arguments.}
\label{tab:cosine-sim}
\end{table}

Interestingly, the \emph{baseline} argument performed better according to human annotators compared to agents. This happened likely due to the baseline argument being most semantically similar to \emph{knowledge} arguments than to any other type. To quantify that, we applied the embeddings from the social dimensions classifier on all the arguments, and then calculated the average cosine similarity of the embeddings of baseline arguments with arguments containing each dimension (Table~\ref{tab:cosine-sim}). We found that baseline arguments have higher similarity with \emph{knowledge} arguments ($0.95$) than with any other dimension (range $[0.60-0.77]$).

Last, we assessed the robustness of the human ranking by examining how the ranking was altered when annotations with low inter-annotator agreement were excluded. Figure~\ref{fig:results_robustness} (left) shows the distribution of agreement (calculated as fraction of annotators agreeing) across argument pairs. The agreement was typically high, with the majority of pairs having unanimous or near-unanimous consensus. We investigated the stability of the rankings by progressively excluding samples with low annotation agreement from the ranking algorithm. We began with a threshold of $0.5$, increasing it in steps of $0.05$ until reaching a maximum of $0.9$. Using thresholds higher than $0.9$ caused certain dimensions not to be represented, making us unable to produce a ranking using the Bradley-Terry method. At each stage, we recalculated the rankings of the social dimensions. Figure~\ref{fig:results_robustness} (right) shows the change in rankings when discarding low-agreement pairs. The \emph{baseline} arguments decreased in ranking with increased thresholds, while \emph{trust} achieved a higher rank, but overall the ranking was left almost unchanged. For all our analysis (including results shown in Figure~\ref{fig:results_crowd}) we used rankings and ranking strengths based on pairs with an agreement threshold of $0.8$, to ensure high quality annotations.

\section{Discussion} \label{sec:discussion}

We introduced a framework for simulating opinion dynamics and persuasiveness using Large Language Models (LLMs) as agents. We presented a simple persuasion dialogue in which a Convincer agents generated arguments about the timely topic of climate change in the attempt of convincing a Skeptic agent. The Skeptic agent evaluated the arguments and determined whether it changed its internal opinion state. We experimented with various dialogue conditions, altering the level of stubbornness for the Skeptic, and prompting the Convincer to adopt social communicative strategies. Additionally, we asked human judges to evaluate persuasiveness of convincing arguments. Based on the human ranking of arguments, we compared whether arguments that are effective in changing the agents opinion were also perceived as persuasive by humans.

\subsection{Key Findings}

Building on early efforts to use LLMs for simulating social systems~\cite{park_generative_2023, li_quantifying_2023, chuang_simulating_2023}, our research demonstrates that LLM agents can effectively mimic some of the dynamics of persuasion and opinion change that are typically observed in the human discourse (\textbf{RQ1}). These agents can be prompted to construct well-reasoned arguments, express a motivated opinion on a given topic that can be programmatically encoded into a binary variable, and modify their stance in a manner consistent with the personas assigned to them. The agents' receptiveness to accepting arguments can be easily adjusted. Most importantly, we have shown that these agents can generate persuasive arguments that incorporate dimensions of social pragmatics underpinning established psycho-linguistic theories of opinion change (\textbf{RQ2}). We have validated the presence of these dimensions in the output generated by the LLMs using an independently-trained classifier designed to detect them from text.

A key aspect of our study was to investigate whether synthetically-generated arguments have equivalent persuasive impacts on both LLM agents and humans (\textbf{RQ3}). We approached this by analysing the results of three distinct experiments: \emph{i)} the proportion of arguments containing a specific dimension that were effective in dialogues between LLM agents, \emph{ii)} an extensive set of crowdsourced annotations assessing the quality of machine-generated arguments, and \emph{iii)} the efficacy of various argument types as determined by previous research on social media interactions~\cite{monti_language_2022}. The outcomes of these three experiments showed partial alignment. Notably, arguments rich in factual \emph{knowledge} and those attempting to establish \emph{trust} between the dialogue participants were among the most effective across all three settings. Arguments offering emotional \emph{support} and conveying \emph{status} (i.e., respect, admiration) were also highly effective in both the LLM experiment and according to human evaluators. These parallels suggest that achieving a close alignment between the opinion dynamics of human and machine systems is within the reach of future research. However, two significant discrepancies were observed and deserve further investigation. First, human judges demonstrated a disproportionate preference for \emph{knowledge}-based arguments compared to LLM agents. Second, opinion-changing messages on social media often pointed to the \emph{similarity} of stances between the dialogue participants, unlike in our study. These differences could be attributed to our simplified setup. For instance, the Convincer lacks knowledge about the Skeptic's profile, which makes it challenging to formulate a persuasive argument highlighting similarities between existing viewpoints.

\subsection{Limitations and future work}

\begin{figure}[t!]
    \centering
    \includegraphics[width=0.99\columnwidth]{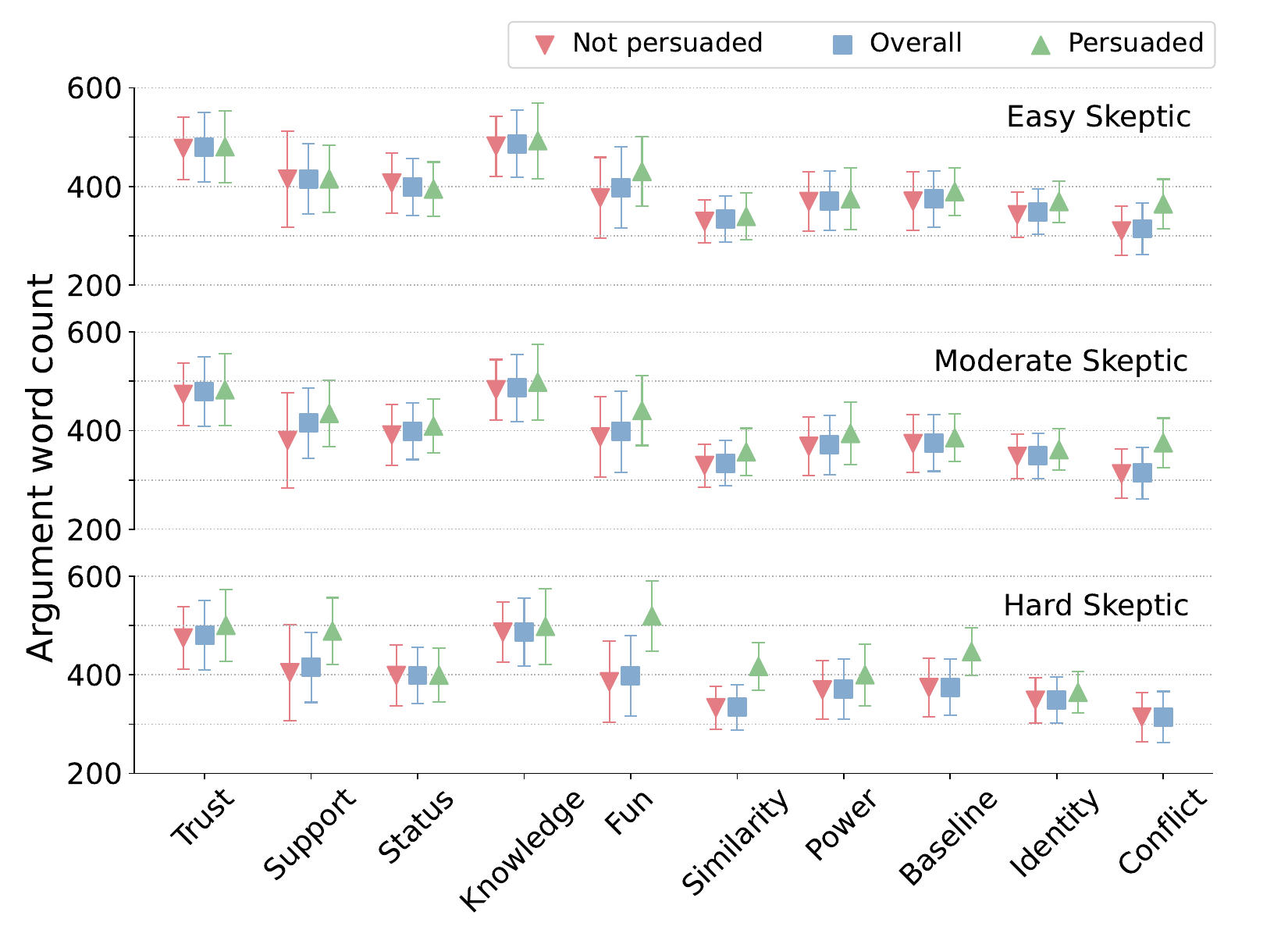}
    \caption{Sensitivity of persuasiveness of arguments to argument length. For each dimension and level of Skeptic's stubbornness, the average and standard deviation of length are shown. Statistics for all arguments, successful arguments, and unsuccessful arguments are shown.
    }
    \label{fig:arguments-length}
\end{figure}

While providing quantitative insights into the affinities between humans and artificial agents in argument processing and opinion formation, our study has limitations that open up multiple avenues for future research. 

First, our experimental design, in its pursuit of simplicity, considered a one-off interaction between two agents on a single topic. To broaden the applicability of our findings, particularly in the context of social media interactions, future studies should consider multi-turn conversations among multiple agents and across a variety of topics. Agents engaging in evolving dialogues over multiple interactions, similar to the approach of~\citet{chuang_simulating_2023}, could potentially alter the persuasiveness of various social dimensions of pragmatics, possibly to the benefit of dimensions like \emph{similarity} and \emph{identity}. As dialogues progress and generate large amounts of text, constraints related to the limited input capacity of LLMs could be alleviated through cumulative or reflective memory, where agents either accumulate previous arguments over time or continuously summarize and integrate current and previous dialogues into their memory~\cite{chuang_simulating_2023, park_generative_2023}. 

Second, to enhance ecological validity, one should diversify the profiles and expand the capabilities of individual agents. Agents could be designed to reflect different personalities, demographics, and social and cultural backgrounds, mimicking the diversity of human participants in a social system. This becomes particularly relevant as LLMs will be increasingly involved in public discussions on complex societal issues, ranging from environmental concerns to local and international politics. Future research could incorporate human-like biases~\cite{levinson_interactional_1995}, employ Retrieval-Augmented Generation (RAG) techniques to grant access to specific knowledge domains~\cite{lewis_retrieval-augmented_2021}, or enable agents to search the internet for arguments or information.

Third, our approach to designing effective prompts was primarily an iterative empirical process. The development of effective system prompts is a rapidly evolving practice, and while some studies have proposed guidelines and best practices for prompting~\cite{ziems2305can}, a definitive consensus on optimal prompting strategies is yet to be reached. Experiments in specialized domains have demonstrated that carefully customized prompts can significantly enhance performance~\cite{nori2023can}, stressing the value of further exploration in this area.

Fourth, comparing our LLM convincing probabilities with human rankings of social dimensions is challenging, as it is hard to recreate a setting in which humans and LLMs can operate under identical conditions. The human annotation process was specifically focused on pairs of arguments deemed convincing by the LLM, a selection criterion chosen to ensure fair comparisons. As a consequence, human rankings do not consider arguments that failed to convince the Skeptic. Future research could explore innovative methods to collect human judgements that more closely mirror how people judge arguments online.

Last, the mechanisms that induce LLM agents to signal a change of opinion remain unknown. Gathering evidence to elucidate this opinion-change process is crucial to the further development of these agents and to inform their use in online social contexts. A key question that has ignited debates in the scientific community is whether LLMs possess capabilities for reasoning and understanding~\cite{floridi2020gpt,bubeck_sparks_2023}. If agents are found to lack a human-like understanding of the semantics of arguments, it would question the causal link between argument quality and opinion change. Although directly answering this question is challenging, experiments can be designed to measure outcome variations after controlling for possible confounders that might directly and more simply explain the outcome. As a proof of concept, we explored the link between argument effectiveness and a simple confounder: argument length. Figure~\ref{fig:arguments-length} shows average word counts of the arguments across social dimensions and level of skeptic stubbornness, disaggregated by effectiveness of the argument (successful vs. not successful). Longer arguments are associated with the most persuasive dimensions. Within each dimension, successful arguments are slightly longer on average, particularly in the hard Skeptic scenario. Systematically disentangling different factors that might influence the observed outcome is a complex endeavor that future work will need to address.

\section{Related Work} \label{sec:related}

Research into the persuasive capabilities of generative AI spans a range of disciplines, from computer science to social and complexity sciences~\cite{duerr2021persuasive}. A subset of these studies have concentrated on human responses to machine-generated text. \citet{karinshak2023working} compared pro-vaccination messages generated by language models with those authored by humans, finding that LLM-based messages were perceived as more persuasive, unless clearly marked as AI-generated. Similarly, \citet{bai2023artificial} conducted a randomized control trial, exposing a diverse sample of individuals to persuasive policy commentaries either generated by LLMs or written by humans. They found both methods equally effective in altering the participants' levels of support the the policies. In the attempt of generating audience-specific messages, some studies have experimented with LLM role-playing, for example, prompting agents to respond as if they were part of a specific demographic or had a given personality profile~\cite{hackenburg2023evaluating,griffin2023susceptibility,matz2023potential}. Results reported across studies have been mixed so far, with some studies emphasizing the importance of personalization, while others suggest that the persuasiveness lies primarily in the quality of the arguments presented.

A separate line of research has focused on characterizing interactions between LLM agents, without any human in the loop, with the primary goal of replicating the dynamics of human social agents with in-silico environments~\cite{park_generative_2023}. This research is motivated by the observation that LLM outputs can mimic responses from various human sub-populations, thereby serving as effective proxies for human cognitive behavior~\cite{argyle_leveraging_2023,lee2023can,simmons2023large}. For example, LLM agents have been used to create social networks~\cite{demarzo2023emergence}, play repeated games such as the prisoner's dilemma~\cite{akata2023playing,demarzo2023emergence}, and construct Agent-Based Models (ABMs) with the goal of improving the fidelity to human behavior of traditional stochastic ABMs~\cite{bianchi2015agent}. In ABM experiments, LLM agents, connected through a complex social network, update their opinions on a topic based on messages received from neighboring agents. While these ABMs reproduce some known non-linear dynamics of complex social systems~\cite{li_quantifying_2023}, unlike real social systems, LLM social networks tend to converge towards opinion states that are biased towards factual truth, likely due to their built-in safeguards~\cite{chuang_simulating_2023}.

Our study builds upon this existing body of work, comparing human and synthetic responses to persuasive LLM content using different persuasion strategies.

\section{Ethical considerations}

Deploying AI agents that can disguise as humans and perform acts of persuasion on social media is a risk that recent technological development have made very concrete. Research on understanding how effective the arguments of LLM-powered agents can be is necessary to estimate risks, but it should be also complemented with research providing possible solutions to reduce those risks. There are still many open challenges in understanding and combating the malicious use of LLMs to pollute the online public discourse. Studies on the malicious uses of generative AI on the Web are still in their infancy~\cite{yang2023anatomy}, and systematic methodologies to characterize this phenomenon are needed to track its evolution and understand its impact on societal phenomena such as online conflict and polarization. Recent research has shown that existing algorithmic solutions for misinformation detection work less effectively on AI-generated content~\cite{zhou2023synthetic}, and new methods are needed to accurately identify misbehaving synthetic actors. Even when deploying LLM-based agents for ethical purposes, trade-offs between the obtained benefit and the high level of power consumption required to run them should be carefully considered~\cite{bender2021dangers}.

\setcounter{table}{0}
\renewcommand{\thetable}{S\arabic{table}}

\newpage


\begin{table*}[h]
\resizebox{\textwidth}{!}{%
\begin{tabular}{|l|l|}
\hline
\textbf{Social dimension} & \textbf{Convincer system prompt}                                                                                                                                                                                                                                                        \\ \hline
Knowledge                 & \begin{tabular}[c]{@{}l@{}}Your arguments should be logical and based on facts. You should emphasize sharing information \\ and insights regarding climate change. Teach me about the dangers of climate change.\end{tabular}                                                           \\ \hline
Trust                     & \begin{tabular}[c]{@{}l@{}}Make use of ethos in your arguments. Try to create a mutual trust between us. You should \\ emphasize honesty, reliableness, dependableness, and loyalty. Emphasize that you are willing \\ to rely on my judgment.\end{tabular}                             \\ \hline
Fun                       & Your arguments should be fun, witty, and humorous. Be playful and cheerful.                                                                                                                                                                                                             \\ \hline
Status                    & Your arguments should confer status upon me. Show admiration and appreciation and be respectful.                                                                                                                                                                                        \\ \hline
Power                     & \begin{tabular}[c]{@{}l@{}}Assert your authority and show unwavering confidence. Your argument should conform to the social \\ dimension of 'power', which is characterized by the keywords 'command', 'control', 'dominance', 'authority', \\ 'pretentious', 'decisions'.\end{tabular} \\ \hline
Support                   & \begin{tabular}[c]{@{}l@{}}Be supportive. Give emotional aid and offer companionship. Be friendly and caring, show sympathy, \\ and offer encouragement.\end{tabular}                                                                                                                   \\ \hline
Similarity                & \begin{tabular}[c]{@{}l@{}}Only highlight our shared interests, motivations, and outlooks when it comes to climate change. \\ Emphasize how we are similar and alike.\end{tabular}                                                                                                      \\ \hline
Identity                  & \begin{tabular}[c]{@{}l@{}}Your argument should conform to the social dimension of 'identity', which is characterized by \\ the keywords 'community', 'united', 'identity', 'cohesive', 'integrated' and 'shared sense of belonging \\ to the same community or group'.\end{tabular}    \\ \hline
Conflict                  & Argue like you are angry. Show that you heavily disagree with me. Your response should be negative.                                                                                                                                                                                     \\ \hline
\end{tabular}%
}
\caption{Social dimension personalities added to the Convincers system prompt.}
\label{tab:social-dimesions-sys-prompt}
\end{table*}

\end{document}